\documentclass[12pt,a4paper]{article}
\usepackage{amsfonts,latexsym}
\usepackage{amsmath,amssymb}
\usepackage{graphicx,color}

\renewcommand{\thefootnote}{\fnsymbol{footnote}}

\begin{document}

\vspace{12mm}

\begin{center}
{{{\Large {\bf Shadow bound of black holes with dark matter halo }}}}\\[10mm]

{Yun Soo Myung\footnote{e-mail address: ysmyung@inje.ac.kr}}\\[8mm]

{Institute of Basic Sciences and Department  of Computer Simulation, Inje University, Gimhae 50834, Korea\\[0pt] }

\end{center}
\vspace{2mm}

\begin{abstract}
We investigate the  shadow cast of black holes immersed in a dark matter halo.
We use the M87* shadow data obtained by the EHT collaboration to constrain two parameters ($M,a$) of dark matter halo  surrounding  a black hole with mass $M_{\rm bh}$.
For $a\ge10.8M$, we find the favored region (shadow bound), while the disfavored region is found for $a<10.8M$ when imposing the EHT results.
This shadow bound is much less  than the observation bound of galaxies ($a\ge 10^4 M$).
\end{abstract}
\vspace{5mm}

\vspace{1.5cm}

\hspace{11.5cm}{Typeset Using \LaTeX}
\newpage
\renewcommand{\thefootnote}{\arabic{footnote}}
\setcounter{footnote}{0}


\section{Introduction}

The significant results of the Event horizon Telescope (EHT) collaboration shed some light on  a new era of black hole (BH) observations.
The image of the M87* BH~\cite{EventHorizonTelescope:2019dse,EventHorizonTelescope:2019ths,EventHorizonTelescope:2019ggy} has inspired many studies on the  BH shadow  to test modified gravity theories.
The recent EHT results focussed on  the center of our galaxy and revealed  curious  images of the  SgrA* BH~\cite{EventHorizonTelescope:2022wkp,EventHorizonTelescope:2022wok,EventHorizonTelescope:2022xqj}.

It is likely  that BHs are not completely isolated objects in the universe.
The shadow of BH with scalar hair was used to test the EHT results~\cite{Khodadi:2020jij}, while the shadow of BHs, worm holes, and naked singularity  obtained from modified gravity theories were employed to constrain their parameters when comparing with the resent EHT results~\cite{Vagnozzi:2022moj}.
There is a strong evidence that dark matter surrounds most galaxies in a halo~\cite{Bertone:2018krk}.
The authors of \cite{Cardoso:2021wlq} have suggested a model  to embed a BH into a dark matter halo.
They used a Hernquist-type density  distribution that are observed in bulges and elliptical galaxies~\cite{Hernquist:1990be}. A lot of recent  studies on this model are found in~\cite{Konoplya:2021ube,Jusufi:2022jxu,Konoplya:2022hbl,Figueiredo:2023gas,Stelea:2023yqo}
and  the effects of dark matter on the shadow of BHs were described  in~\cite{Xavier:2023exm}.

In this work, we will study the  shadow of BHs immersed  in a dark matter halo.
We use the M87* shadow data obtained by the EHT collaboration to constrain  two parameters ($M,a$) of dark matter halo including  a BH with mass with $M_{\rm bh}$ at the center.
We analyze and discuss the photon ring and shadow radius (critical impact parameter).
For $a\ge10.8M$, we find the favored  region (shadow bound), while the disfavored region is found for $a<10.8M$ when imposing the EHT results.
If one uses the SgrA* shadow data to constrain two parameters, the shadow bound is given by $a\ge 18M$ for Keck Observatory and $a\ge 50M$ for the Very Large Telescope Interferometer.
These shadow bounds are  smaller  than the observation bound of galaxies ($a\ge 10^4 M$).

\section{BHs with dark matter halo}
First of all, we would like to introduce  the Hernquist-type density  distribution~\cite{Hernquist:1990be}
\begin{equation}
\rho_{\rm H}(r)=\frac{Ma}{2\pi r(r+a)^3},
\label{H-den}
\end{equation}
where $M$ is the total  mass of the galactic halo and $a$ denotes a typical length-scale of the galaxy.
Including a BH with mass $M_{\rm bh}$ at the center, the resulting spacetime is described by~\cite{Cardoso:2021wlq}
\begin{equation}
ds^2=-f(r)dt^2+\frac{dr^2}{1-\frac{2m(r)}{r}}+r^2d\Omega^2, \label{sss}
\end{equation}
where the metric function $f(r)$ could be derived from the mass function,
\begin{equation}
m(r)=M_{\rm bh}+\frac{Mr^2}{(r+a)^2}\Big(1-\frac{2M_{\rm bh}}{r}\Big)^2.
\end{equation}
This mass function goes over to the Hernquist density (\ref{H-den}) at large scales, whereas it describes a BH with mass $M_{\rm bh}$ at small scales.
Solving the Einstein equation  with asymptotically flat spacetime
\begin{equation}
\frac{rf'}{2f}=\frac{m(r)}{r-2m(r)},
\end{equation}
the metric function $f(r)$ is obtained analytically  as
\begin{eqnarray}
f(r)=\Big(1-\frac{2M_{\rm bh}}{r}\Big)e^{\Upsilon}
\end{eqnarray}
with
\begin{equation}
\Upsilon(r)=\sqrt{\frac{M}{2a-M+4M_{\rm bh}}}\Bigg[-\pi +2\arctan\Big[\frac{r+a-M}{\sqrt{M(2a-M+4M_{\rm bh})}}\Big]\Bigg].
\end{equation}
Here, $e^{\Upsilon(r)}$ is regarded as a redshift factor. We note that the event horizon is located at $r_+=2M_{\rm bh}$ as for the Schwarzschild solution and the  ADM mass of spacetime  takes the form of $M+M_{\rm bh}$.
This solution could be regarded as a model for a supermassive BH in the center of a galaxy surrounded by a dark matter halo.
To mimic observations of galaxies, one requires that $a \geq10^4M$ and a hierarchy of scales: $M_{\rm bh} \ll M \ll a$. The compactness of dark matter halo is measured by a quantity of ${\cal C}=M/a$ with $G=c=1$ unites.
The matter density associated with this spacetime is changed  by
\begin{equation}
\rho(r)=\frac{M(a+2M_{\rm bh})}{2\pi r(a+r)^3}\Big(1-\frac{2M_{\rm bh}}{r}\Big),
\end{equation}
where one recovers $\rho_{\rm H}(r)$ at large distances and for $a\gg M_{\rm bh}$.

\section{Photon rings and shadow radius}
\begin{figure*}[t!]
   \centering
   \includegraphics{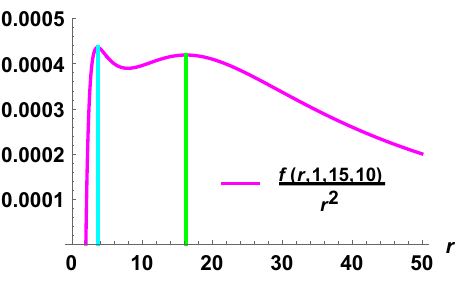}
 \hfill%
    \includegraphics{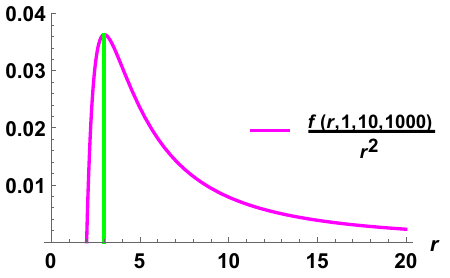}
\caption{(Left) Effective potential $V(r)$ as function of $r\in[r_+=2,50]$  for $M_{\rm bh}=1,~M=15,~a=10$ with ${\cal C}=1.5$.  Two unstable photon rings emerge at $r_{\gamma,1}=16.33$ and  $r_{\gamma,3}=3.68$.  (Right) Effective potential $V(r)$ as function of $r\in[r_+=2,20]$ for $M_{\rm bh}=1,~M=10,~a=1000$ with ${\cal C}=10^{-4}$. Single unstable photon ring appears at $r_{\gamma,1}=3.00$. }
\end{figure*}
Firstly, we wish to derive the photon ring radius $r_\gamma$.
For this purpose, the effective potential for null geodesics is given by~\cite{Xavier:2023exm}
\begin{equation}
V(r)=\frac{f(r)}{r^2}.
\end{equation}
A photon ring  corresponds to a critical point of $V(r)$, that is,
\begin{equation}
V'(r)|_{r=r_\gamma}=0. \label{pot-d}
\end{equation}
On the other hand, one has exact photon ring radii  by finding  three roots of $r=3m(r)$~\cite{Cardoso:2021wlq}
\begin{equation}
r_{\gamma,i}(M_{\rm bh},M,a),\quad  {\rm for} \quad  i=1,2,3,
\end{equation}
whose expressions are too complicated to write down here.
For $a\gg M$, however,  one keeps up to ${\cal O}(1/a^3)$ to obtain three approximate roots
\begin{eqnarray}
r_{\gamma,a1}(M_{\rm bh},M,a)&=&3M_{\rm bh}\Big(1+\frac{MM_{\rm bh}}{a^2}\Big), \label{gam-1} \\
r_{\gamma,a2}(M_{\rm bh},M,a)&=&\frac{3M}{2}-a-\frac{3M M_{\rm bh}^2}{2a^2}+i\sqrt{3aM}\Big(1+\frac{3M^2-8MM_{\rm bh}-16M^2_{\rm bh}}{128a^2}\Big), \label{gam-2} \\
r_{\gamma,a3}(M_{\rm bh},M,a)&=&\frac{3M}{2}-a-\frac{3M M_{\rm bh}^2}{2a^2}-i\sqrt{3aM}\Big(1+\frac{3M^2-8MM_{\rm bh}-16M^2_{\rm bh}}{128a^2}\Big), \label{gam-3}
\end{eqnarray}
where the last two become complex conjugate variables.

As is shown in Fig. 1a with $M_{\rm bh}=1,M=15,a=10$~\cite{Guo:2022ghl,Xavier:2023exm} where the dark matter distribution concentrates near the horizon of BH with ${\cal C}=1.5$, three photon rings determined by Eq.(\ref{pot-d}) are the same as those ($r_{\gamma,i}=16.33,~8,~3.68$)  determined by $r=3m(r)$. However, they are quite different from approximate roots ($r_{\gamma,ai}=3.45,~12.41+25.54i,~12.41-25.54i$).  From Fig. 1b with $M_{\rm bh}=1,M=10,a=1000$ whose dark matter compactness is very small (${\cal C}=10^{-4}$), we find that single photon ring ($r_\gamma=3$) determined by Eq.(\ref{pot-d}) is  the same as the first real one of  ($r_{\gamma,i}=3,~-985+172.64i,~-985-172.64i$)  determined by $r=3m(r)$. These exact photon ring radii  are nearly  the same as  approximate roots ($r_{\gamma,ai}=3,~-985+173.2i,~-985-173.2i$). It is worth to note that two complex roots are not real photon ring radii for  ${\cal C}=10^{-4}$, but
they turned out to be real for  ${\cal C}=1.5$. This is a significant difference between  $M_{\rm bh}=1,M=15,a=10$ (two unstable photon rings) and  $M_{\rm bh}=1,M=10,a=1000$ (single unstable photon ring).
These are related to an appearance of two solutions of $r_{2,3}=M-a\pm \sqrt{M^2-2Ma -4MM_{\rm bh}}$ to $r=2m(r)$, in addition to $r_+=2M_{\rm bh}$.
Even though  two unstable photon rings for $M_{\rm bh}=1,M=15,a=10$ may show theoretical interest, they are not legitimate to describe astrophysical BH-halo system.
Hence, we wish to discard the case of $M_{\rm bh}=1,M=15,a=10$ for computing critical impact parameter.

In Fig. 2a, we display  photon ring radii  $r_{\gamma,a1}$ and $r_{\gamma,a1}$ as functions of $a$ for  $M_{\rm bh}=1,~M=15$. Two approach $3M_{\rm bh}$ (Schwarzschild case) for $a>M$, but they are different for $a<M$ ($r_{\gamma,1}$ has a discontinuity at $M=10$). This implies that the approximate root $r_{\gamma,a1}$ could describe photon ring radius only  for $a>M$.
\begin{figure*}[t!]
   \centering
   \includegraphics{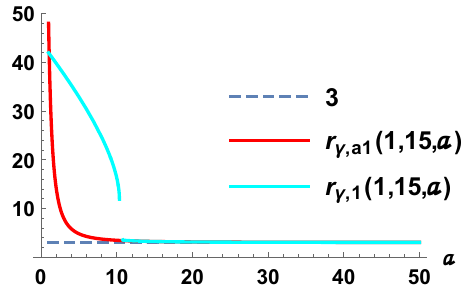}
 \hfill%
    \includegraphics{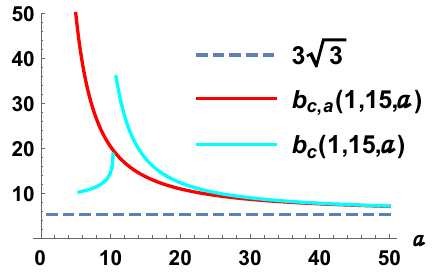}
\caption{(Left) Photon ring radii  $r_{\gamma,a1}$ and $r_{\gamma,1}$ as functions of $a\in[1,50]$ for  $M_{\rm bh}=1,~M=15$. Two approach $3M_{\rm bh}$ (=3, Schwarzschild case) for $a>M$. (Right)  Critical impact parameters $b_{c,a}$ and $b_c$ as functions of $a\in[1,50]$ for  $M_{\rm bh}=1,~M=15$. Two parameters approach $3\sqrt{3}M_{\rm bh}$ (=5.196, Schwarzschild case) for $a>M$. }
\end{figure*}
\begin{figure*}[t!]
   \centering
   \includegraphics{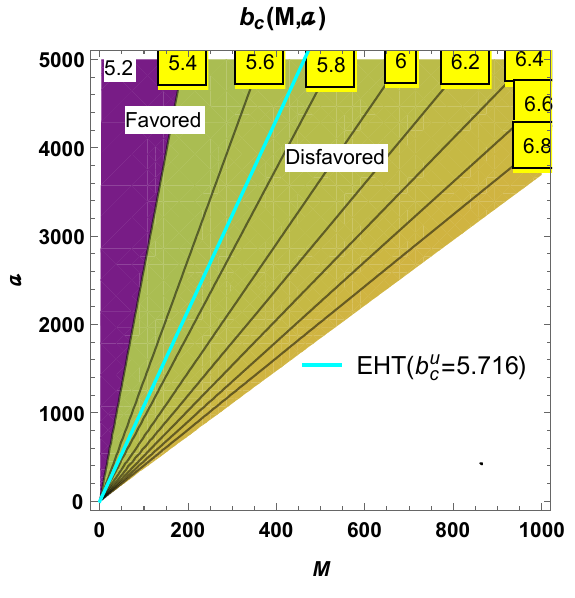}
\caption{Contour plot of exact critical impact parameter $b_c$ as functions of $M\in[1,1000]$ and $a\in[0,5000]$ for $M_{\rm bh}=1$. One has $b_c(10,5000)\simeq5.2$.  Different lines denote the lines for the same impact parameters.  Based on the EHT results, the favored region ($5.2\le b_c\le 5.716$) is separated from the disfavored region ($b_c>5.716$)  by the cyan line ($b_c^u=5.716$). This line could be represented precisely by $a(M)=10.8M$ which divides $a>10.8M$ (favored region) and $a<10.8M$ (disfavored region). }
\end{figure*}
Now, we have a position to find the shadow radius to test the EHT results.
The shadow radius is defined by the critical impact parameter because of spherically symmetric spacetime Eq.(\ref{sss})
\begin{equation}
r_{\rm sh}\to b_c=\frac{r}{\sqrt{f(r)}}\mid_{r=r_\gamma}.\label{r-cri}
\end{equation}
We have two types of critical impact parameters as
\begin{eqnarray}
b_{c,a}(M_{\rm bh},M,a)&=&3\sqrt{3} M_{\rm bh}\Big[1+\frac{M}{a}+\frac{M(5M-18M_{\rm bh})}{6a^2}\Big], \label{abc} \\
b_{c}(M_{\rm bh},M,a)&=&3\sqrt{3} M_{\rm bh}e^{-0.5\Upsilon(r_{\gamma,1})}, \label{bc}
\end{eqnarray}
where the former is an approximate one~\cite{Cardoso:2021wlq} obtained  by keeping up to ${\cal O}(1/a^3)$ when substituting $r_{\gamma,a1}(M_{\rm bh},M,a)$ to Eq.(\ref{r-cri}), while the latter is an exact one found by plugging $r_{\gamma,1}(M_{\rm bh},M,a)$ into  Eq.(\ref{r-cri}) even though its form is complicated. In deriving the latter, we use $r/\sqrt{1-2M_{\rm bh}/r}|_{r=r_{\gamma,1}}\simeq3\sqrt{3}M_{\rm bh}$ [keeping up to ${\cal O}(1/a^4)$].
We check that $\lim_{M\to 0}b_c=\lim_{a\to\infty}b_c=3\sqrt{3}M_{\rm bh} (=b_{c,S}$, Schwarzschild case) as is easily  shown by $b_{c,a}$.
As is shown in Fig. 2b, two critical impact parameters  approach $3\sqrt{3}M_{\rm bh}$  for $a>M$, whereas two are different for $a<M$. So, $b_{c,a}$  could be used  to describe shadow radius only for $a>M$. On the other hand, the critical impact parameter $b_c$ has a discontinuity at $M\simeq11$ because the photon ring radius $r_{\gamma,1}$  has a discontinuity at $M=10$.

\section{Shadow bounds}
\begin{figure*}[t!]
   \centering
   \includegraphics{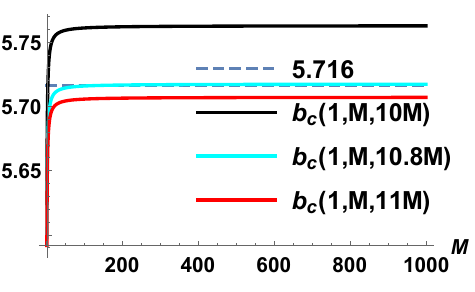}
 \hfill%
    \includegraphics{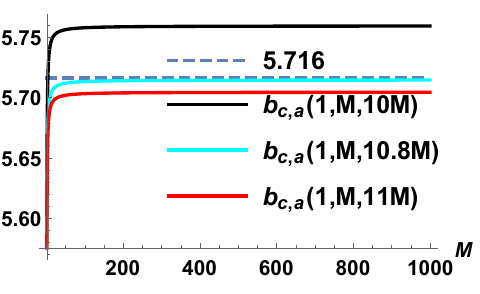}
\caption{(Left) Three impact parameters $b_c$ as functions of $M\in[1,1000]$ with $a=10M,~10.8M,~11M$. The case of $a=10.8M$ agrees with  $b_c^u=5.716$. (Right) Three impact parameters $b_{c,a}$ as functions of $M\in[1,1000]$ with $a=10M,~10.8M,~11M$. The case of $a=10.8M$  is consistent with   $b^u_{c,a}=5.716$. }
\end{figure*}

In Fig. 3, we show the  favored (disfavored) region by making use of the results  of the EHT collaboration (M87*)~\cite{EventHorizonTelescope:2019dse}. If the relative deviation from the Schwarzschild  result ($b_{c,S}\simeq 5.196$) is less (greater) than 10\%, the solution is in the favorable (disfavored) region~\cite{Xavier:2023exm}. The line of $b_c^u=5.716$ represents the upper limit and it is also given by $a(M)=10.8M$ (see Fig. 4a). Therefore, the favored region (shadow bound) is represented by $a\ge10.8M$. For the SgrA*~\cite{EventHorizonTelescope:2022xqj}, the favored region is represented by $a\ge 18M(b_c^u=5.5)$ for Keck Observatory and $a\ge 50M(b_c^u=5.3)$ for the Very Large Telescope Interferometer.  We note that the observation of galaxies corresponds to the regime of $a\ge 10^4M$.

Similarly, we can show the region favored (disfavored) when adopting the approximate critical impact parameter $b_{c,a}$. If the relative deviation from the Schwarzschild  result is less (greater) than 10\%, the solution is in the favorable (disfavored) region.  Here, the line of $b_{c,a}^u=5.716$ represents the upper limit and it is also given by $a(M)=10.8M$ (see Fig. 4b).  Thus, one finds that  the favored (shadow) region is represented by the same line of $a\ge10.8M$.

\section{Conclusions}

We have studied  the  shadow cast of  black holes immersed  in a dark matter halo.
We have used the M87* shadow data obtained by the EHT collaboration to constrain  two parameters ($M,a$) of dark matter halo  surrounding  a black hole mass with $M_{\rm bh}$.
We analyzed and discussed the photon ring and shadow radius (critical impact parameter).
For a case of  ($M_{\rm bh}=1,M=15,a=10$)~\cite{Guo:2022ghl,Xavier:2023exm} where the dark matter distribution concentrates near the horizon of BH with ${\cal C}=1.5$, we have three photon rings ($r_{\gamma,i}=16.33,~8,~3.68$).
For a case of  ($M_{\rm bh}=1,M=10,a=1000$) whose dark matter compactness is very small like as ${\cal C}=10^{-4}$, we find  single photon ring ($r_{\gamma,1}=3$) with two complex rings ($r_{\gamma,2}=-985+172.64 i,r_{\gamma,3}=-985-172.64 i$).
Here, we observe that the last two real photon rings for ${\cal C}=1.5$  become  complex quantities  for ${\cal C}=10^{-4}$.  Also, the former is not suitable for describing astrophysical BH-halo systems and thus, this case  was discarded from the consideration.

The shadow cast is described by the critical impact parameter $b_c$.
If the relative deviation from the Schwarzschild  result ($b_{c,S}=3\sqrt{3}= 5.196$) is less (greater) than 10\%, the solution is in the favorable (disfavored) region~\cite{Xavier:2023exm}. The line of $b_c^u=5.716$ represents the upper limit and it is also given by $a(M)=10.8M$ (see Fig. 4a). Therefore, the favored region (shadow bound) is represented by $a\ge10.8M$. For the SgrA*~\cite{EventHorizonTelescope:2022xqj}, the shadow bound  is represented by $a\ge 18M(b_c^u=5.5)$ for Keck Observatory and $a\ge 50M(b_c^u=5.3)$ for the Very Large Telescope Interferometer. We have the same result when using the approximate shadow radius $b_{c,a}$.  It is worth noting that the observation of galaxies corresponds to the regime of $a\ge 10^4M$. Hence, these shadow bounds are  smaller  than the observation bound of galaxies.


\newpage

\begin{thebibliography}{99}
\bibitem{EventHorizonTelescope:2019dse}
K.~Akiyama \textit{et al.} [Event Horizon Telescope],
Astrophys. J. Lett. \textbf{875}, L1 (2019)
doi:10.3847/2041-8213/ab0ec7
[arXiv:1906.11238 [astro-ph.GA]].

\bibitem{EventHorizonTelescope:2019ths}
K.~Akiyama \textit{et al.} [Event Horizon Telescope],
Astrophys. J. Lett. \textbf{875}, no.1, L4 (2019)
doi:10.3847/2041-8213/ab0e85
[arXiv:1906.11241 [astro-ph.GA]].

\bibitem{EventHorizonTelescope:2019ggy}
K.~Akiyama \textit{et al.} [Event Horizon Telescope],
Astrophys. J. Lett. \textbf{875}, no.1, L6 (2019)
doi:10.3847/2041-8213/ab1141
[arXiv:1906.11243 [astro-ph.GA]].

\bibitem{Khodadi:2020jij}
M.~Khodadi, A.~Allahyari, S.~Vagnozzi and D.~F.~Mota,
JCAP \textbf{09}, 026 (2020)
doi:10.1088/1475-7516/2020/09/026
[arXiv:2005.05992 [gr-qc]].



\bibitem{EventHorizonTelescope:2022wkp}
K.~Akiyama \textit{et al.} [Event Horizon Telescope],
Astrophys. J. Lett. \textbf{930}, no.2, L12 (2022)
doi:10.3847/2041-8213/ac6674
[arXiv:2311.08680 [astro-ph.HE]].

\bibitem{EventHorizonTelescope:2022wok}
K.~Akiyama \textit{et al.} [Event Horizon Telescope],
Astrophys. J. Lett. \textbf{930}, no.2, L14 (2022)
doi:10.3847/2041-8213/ac6429
[arXiv:2311.09479 [astro-ph.HE]].

\bibitem{EventHorizonTelescope:2022xqj}
K.~Akiyama \textit{et al.} [Event Horizon Telescope],
Astrophys. J. Lett. \textbf{930}, no.2, L17 (2022)
doi:10.3847/2041-8213/ac6756
[arXiv:2311.09484 [astro-ph.HE]].

\bibitem{Khodadi:2020jij}
M.~Khodadi, A.~Allahyari, S.~Vagnozzi and D.~F.~Mota,
JCAP \textbf{09}, 026 (2020)
doi:10.1088/1475-7516/2020/09/026
[arXiv:2005.05992 [gr-qc]].


\bibitem{Vagnozzi:2022moj}
S.~Vagnozzi, R.~Roy, Y.~D.~Tsai, L.~Visinelli, M.~Afrin, A.~Allahyari, P.~Bambhaniya, D.~Dey, S.~G.~Ghosh and P.~S.~Joshi, \textit{et al.}
Class. Quant. Grav. \textbf{40}, no.16, 165007 (2023)
doi:10.1088/1361-6382/acd97b
[arXiv:2205.07787 [gr-qc]].


\bibitem{Bertone:2018krk}
G.~Bertone and T.~Tait, M.P.,
Nature \textbf{562}, no.7725, 51-56 (2018)
doi:10.1038/s41586-018-0542-z
[arXiv:1810.01668 [astro-ph.CO]].

\bibitem{Cardoso:2021wlq}
V.~Cardoso, K.~Destounis, F.~Duque, R.~P.~Macedo and A.~Maselli,
Phys. Rev. D \textbf{105}, no.6, L061501 (2022)
doi:10.1103/PhysRevD.105.L061501
[arXiv:2109.00005 [gr-qc]].

\bibitem{Hernquist:1990be}
L.~Hernquist,
Astrophys. J. \textbf{356}, 359 (1990)
doi:10.1086/168845

\bibitem{Konoplya:2021ube}
R.~A.~Konoplya,
Phys. Lett. B \textbf{823}, 136734 (2021)
doi:10.1016/j.physletb.2021.136734
[arXiv:2109.01640 [gr-qc]].

\bibitem{Jusufi:2022jxu}
K.~Jusufi,
Eur. Phys. J. C \textbf{83}, no.2, 103 (2023)
doi:10.1140/epjc/s10052-023-11264-w
[arXiv:2202.00010 [gr-qc]].

\bibitem{Konoplya:2022hbl}
R.~A.~Konoplya and A.~Zhidenko,
Astrophys. J. \textbf{933}, no.2, 166 (2022)
doi:10.3847/1538-4357/ac76bc
[arXiv:2202.02205 [gr-qc]].

\bibitem{Figueiredo:2023gas}
E.~Figueiredo, A.~Maselli and V.~Cardoso,
Phys. Rev. D \textbf{107}, no.10, 104033 (2023)
doi:10.1103/PhysRevD.107.104033
[arXiv:2303.08183 [gr-qc]].

\bibitem{Stelea:2023yqo}
C.~Stelea, M.~A.~Dariescu and C.~Dariescu,
Phys. Lett. B \textbf{847}, 138275 (2023)
doi:10.1016/j.physletb.2023.138275
[arXiv:2309.13651 [gr-qc]].

\bibitem{Xavier:2023exm}
S.~V.~M.~C.~B.~Xavier, H.~C.~D.~Lima, Junior. and L.~C.~B.~Crispino,
Phys. Rev. D \textbf{107}, no.6, 064040 (2023)
doi:10.1103/PhysRevD.107.064040
[arXiv:2303.17666 [gr-qc]].

\bibitem{Guo:2022ghl}
G.~Guo, Y.~Lu, P.~Wang, H.~Wu and H.~Yang,
Phys. Rev. D \textbf{107}, no.12, 124037 (2023)
doi:10.1103/PhysRevD.107.124037
[arXiv:2212.12901 [gr-qc]].
\end{thebibliography}
\end{document}